\documentclass{aa} % for the letters 

\usepackage{natbib,twoopt}
\usepackage[breaklinks=true]{hyperref} %% to avoid \citeads line fills
\bibpunct{(}{)}{;}{a}{}{,}             %% natbib format for A&A and ApJ
\makeatletter
  \newcommandtwoopt{\citeads}[3][][]{\href{http://adsabs.harvard.edu/abs/#3}%
    {\def\hyper@linkstart##1##2{}%
     \let\hyper@linkend\@empty\citealp[#1][#2]{#3}}}
  \newcommandtwoopt{\citepads}[3][][]{\href{http://adsabs.harvard.edu/abs/#3}%
    {\def\hyper@linkstart##1##2{}%
     \let\hyper@linkend\@empty\citep[#1][#2]{#3}}}
  \newcommandtwoopt{\citetads}[3][][]{\href{http://adsabs.harvard.edu/abs/#3}%
    {\def\hyper@linkstart##1##2{}%
     \let\hyper@linkend\@empty\citet[#1][#2]{#3}}}
  \newcommandtwoopt{\citeyearads}[3][][]%
    {\href{http://adsabs.harvard.edu/abs/#3}
    {\def\hyper@linkstart##1##2{}%
     \let\hyper@linkend\@empty\citeyear[#1][#2]{#3}}}
\makeatother

\usepackage{graphicx}
\usepackage{txfonts}
\addto\extrasenglish{
    
}
\begin{document}

\title{A Multiwavelength Study of ELAN Environments (AMUSE$^2$)}
\subtitle{Ubiquitous dusty star-forming galaxies associated with enormous Ly$\alpha$ nebulae on megaparsec scales}

\author{Marta Nowotka\inst{1}
        \and
        Chian-Chou Chen\inst{2}
        \and
        Fabrizio Arrigoni Battaia\inst{3}
        \and
        Michele Fumagalli\inst{4,5,6}
        \and
        Zheng Cai \inst{7}
        \and
        Elisabeta Lusso\inst{8,9}
        \and
        J. Xavier Prochaska\inst{10,11}
        \and
        Yujin Yang\inst{12}
        }
        
\institute{
Department of Physics, Colorado College, 14 E. Cache La Poudre Street, Colorado Springs, CO 80903, USA 
\and
Academia Sinica Institute of Astronomy and Astrophysics (ASIAA), No. 1, Sec. 4, Roosevelt Rd., Taipei 10617, Taiwan \\ \email{ccchen@asiaa.sinica.edu.tw}
\and
Max-Planck-Institut fur Astrophysik, Karl-Schwarzschild-Str 1, D-85748 Garching bei M\"{u}nchen, Germany% M\"{u}̈nche
\and
Dipartimento di Fisica G. Occhialini, Università degli Studi di Milano Bicocca, Piazza della Scienza 3, 20126 Milano, Italy
\and
Centre for Extragalactic Astronomy, Durham University, South Road, Durham DH1 3LE, UK
\and
Institute for Computational Cosmology, Durham University, South Road, Durham DH1 3LE, UK
\and
Department of Astronomy, Tsinghua University, Beijing 100084, China
\and
Dipartimento di Fisica e Astronomia, Università di Firenze, Via G. Sansone 1, 50019 Sesto Fiorentino, Firenze, Italy
\and
INAF – Osservatorio Astrofisico di Arcetri, 50125 Florence, Italy
\and
UCO/Lick Observatory, University of California 2013 Santa Cruz, Santa Cruz, CA 95064, USA
\and
Kavli Institute for the Physics and Mathematics of the Universe (Kavli IPMU), 5-1-5 Kashiwanoha, Kashiwa, 277-8583, Japan
\and
Korea Astronomy and Space Science Institute, 776 Daedeokdae-ro, Yuseong-gu, Daejeon 34055, Republic of Korea
}

\date{To be submitted to A\&A}

\abstract{We have been undertaking a systematic survey at 850\,$\mu$m based on a sample of four prototypical $z\sim2-3$ enormous Ly$\alpha$ nebulae (ELANe) as well as their megaparsec-scale (Mpc-scale) environments to study the physical connections between ELANe and their coeval dusty submillimeter galaxies (SMGs). By analysing the SCUBA-2 data with self-consistent Monte Carlo simulations to construct the number counts, here, we report on the overabundance of 850\,$\mu$m-selected submillimeter sources around all the four ELANe, by a factor of 3.6$\pm$0.6 (weighted average) compared to the blank fields. This suggests that the excessive number of submillimeter sources are likely to be part of the Mpc-scale environment around the ELANe, corroborating the co-evolution scenario for SMGs and quasars; this is a process which may be more commonly observed in the ELAN fields. If the current form of the underlying count models continues toward the fainter end, our results would suggest an excess of the 850\,$\mu$m extragalactic background light by a factor of between 2-10, an indication of significant background light fluctuations on the survey scales. Finally, by assuming that all the excessive submillimeter sources are associated with their corresponding ELAN environments, we estimate the SFR densities of each ELAN field, as well as a weighted average of $\Sigma$SFR=1200$\pm$300\,$M_\odot$\,yr$^{-1}$\,Mpc$^{-3}$, consistent with that found in the vicinity of other quasar systems or proto-clusters at similar redshifts; {in addition, it is a factor of about 300 greater than the cosmic mean. } 
}

\titlerunning{AMUSE$^2$ II: DSFG around ELAN}
\authorrunning{Nowotka, Chen, Arrigoni Battaia et al.} 
\maketitle

\section{Introduction} \label{sec:intro}
Galaxies that are selected at submillimeter wavelengths, in particular at 850 micron, contain sites that exhibit some of the most intensive star formation across the Universe, with estimated star-formation rates up to thousands of solar masses per year (e.g., \citealt{Barger:2012aa,Swinbank2014,Cowie2017,Ugne2020}). They are predominantly located at $z=1-3,$ with a long tail toward $z>4$ (\citealt{Barger2000,Chapman2005,Wardlow2011,Simpson2014,TC2016,Stach2018,Ugne2020,Smail2020}), matching the trend of cosmic star formation rate (SFR) densities (\citealt{MD2014}). Evidence that has been collected since the discovery of the first submillimeter galaxies (SMGs; \citealt{Smail1997,Barger1998,Hughes1998}) points to an evolutionary scenario whereby these massive (10$^{10}$-10$^{11}$~M$_{\odot}$ in stellar masses; \citealt{daCunha2015,Michalowski2017}) dusty star-forming galaxies would evolve into the likes of local massive elliptical galaxies, undergoing subsequent multiple phase transitions that include optically bright quasars, compact quiescent galaxies, and dry mergers (e.g., \citealt{Toft2014}). 

The latest evidence supporting this hypothetical evolution comes from galaxy clustering analyses, which have been enabled thanks to the sufficiently large-scale submillimeter surveys that have only recently become available (e.g., \citealt{Geach2017}). These studies have found, via the measurements of both auto-correlation functions and cross-correlation functions, that SMGs are predominantly located in dark-matter halos with masses of around 10$^{12}$-10$^{13}$~M$_{\odot}$ at $z\sim1-3$ (\citealt{Hickox2011,TC2016,TC2016b,Wilkinson2017,An2019,Garcia-Vergara2020,Lim2020}). This is consistent with their being the predicted progenitors of local massive elliptical galaxies, as is the case for the bright quasars at similar redshifts (\citealt{White2012,Timlin2018}, and references therein.)

In general, subsequent active black hole accretion that coincides with or follows intensive star-forming events has long been proposed, even in the local Universe (e.g., \citealt{Hopkins2006}), to partly explain the rapid quenching of star formation through feedback. {This hypothesis, together with the clustering analyses, would suggest an intimate co-evolution between black hole growth and star formation, such that detecting star formation in and around quasar host galaxies is expected. Indeed, the correlated signals found via cross-correlation analyses between infrared background measurements and quasar samples have further supported, on average, 
%further evidence of, on average, 
the co-existing nature of the two populations on halo scales \citep{Wang:2015aa}.} In addition, targeted submillimeter observations on well-defined AGN samples have revealed overdensities of submillimeter sources, as indicated by excessive submillimeter source counts, around some high-redshift radio galaxies (e.g., \citealt{Stevens2003,Dannerbauer2014}), as well as IR luminous AGN selected by the WISE satellite (e.g., \citealt{Jones2017}), again suggesting the possible co-evolution and co-existence of both active black hole accretion and dusty star formation on Mpc scales around massive halos. {However, these studies have often reported overdensity only in a few fields over a larger parent sample, casting the question that the co-evolution may not be universal and can only occur under certain conditions or preferentially around a certain type of quasar \citep{Rigby:2014aa,Zeballos:2018aa}}.

\begin{figure*}[ht]
\centering
\includegraphics[scale = 0.4]{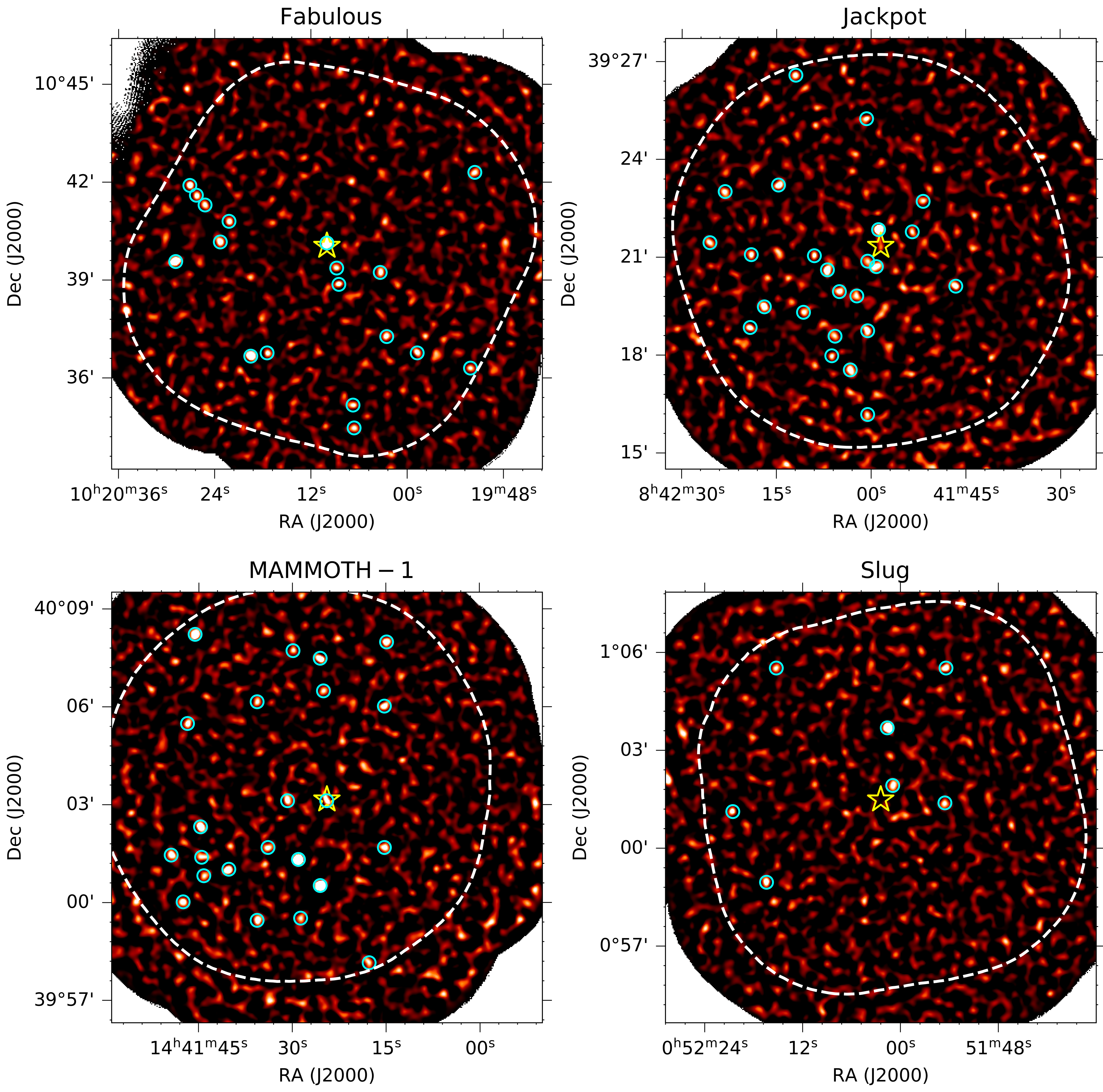}
\caption{SCUBA-2 signal-to-noise maps at 850 $\mu$m for the Fabulous, Jackpot, MAMMOTH-1, and Slug ELANe (yellow star). Sources detected at over 4\,$\sigma$ are marked by cyan circles, and the size of the circles represents $2\times$ the FWHM of the beam. The white curve bounds the effective area of every map, which is about 10$'$ {($\sim$5\,Mpc) in diameter}. The quasars that contribute significantly to the powering of the corresponding ELAN are marked as yellow stars (coordinates in \autoref{table:params}).}
\label{fig:det}
\end{figure*} 

To improve our understanding of this co-evolution {and to expand the quasar types studied}, as part of the A MUltiwavelngth Study of ELAN Environment (AMUSE$^2$; \citealt{Arrigoni-Battaia:2021aa,Chen:2021aa}), we have been conducting submillimeter observations around Enormous Lyman Alpha nebulae (ELANe). ELANe represent the extreme of known Ly$\alpha$ nebulae
%one of the densest structures at $z=2-3$, 
with exceptionally bright (SB$_{\rm Ly\alpha} \sim 10^{-17}$~erg~s$^{-1}$~cm$^{-2}$~arcsec$^{-2}$) Ly$\alpha$ emission over $>100$\,kpc in physical scales, normally embodying multiple powering sources such as AGN, quasars, and other galaxy populations. Current statistics show that only $\sim4\%$ of bright quasars are associated with ELANe, translating into a number density of few times $10^{-6}$~Mpc$^{-3}$ (\citealt{FAB2019,Cai:2019aa}; Arrigoni Battaia et al. submitted). {Spectral analyses focusing on the powering of ELANe have suggested a large amount of cool and dense gas ($\gtrsim10^{10}$M$_\odot$ and $\gtrsim$1\,cm$^{-3}$; \citealt{Fab2015,Hennawi2015}), a finding that is further strengthened by the detection of molecular gas via CO in the circumgalactic medium (CGM) around a central group of galaxies in the MAMMOTH-I ELAN \citep{Emonts:2019aa}. ELANe therefore represent one of the ideal quasar samples for a comprehensive study of the co-evolution between SMGs and quasars and extending the studies to one of the rarest and likely densest regimes.}
%Models predict these structures only occur one in 10$^7$ halos (? and references), 

Our first results intriguingly show a factor of four in overdensities for submillimeter sources around one ELAN, MAMMOTH-1 ($z=2.32$; \citealt{Cai2017,FAB2018b}). In this {paper}, we report our 
%full 
results on three more ELANe, the Jackpot nebula ($z=2.04$; \citealt{Hennawi2015}), the Slug nebula ($z=2.28$; \citealt{Cantalupo14}), and the Fabulous nebula ($z=3.17$; \citealt{FAB2018a}), %representing the %majority of the currently known  
{the first discovered ELANe that marked the start of studies regarding this new class of Lyman alpha nebulae.} %(\citealt{Decarli2020}). %(see Arrigoni Battaia et al. submitted). 
We present our data and analyses in Sections 2 and 3, along with a summary and discussion of the impact of our results in Section 4. We assume the cosmological parameters $H_0=70$\,km\,s$^{-1}$\,Mpc$^{-1}$, $\Omega_{\rm M}=0.3,$ and $\Omega_\Lambda=0.7$. In this cosmology, 1$''$ corresponds to about 7.6 -- 8.4 physical kpc at the redshift range of our sample (Table \ref{table:params}). All distances reported in this work are proper. {Finally, in this paper we define overdensity as $n_{\rm src}/n_{\rm field}$, in which $n_{\rm src}$ is the number density of the submillimeter sources in the ELAN fields and $n_{\rm field}$ is that for the blank fields.}

\section{Observations and data reduction} 
\label{sec:observ}

The dataset for the four ELAN fields studied in this work were acquired with the Submillimetre Common-User Bolometer Array 2 (SCUBA-2; \citealt{Holland2013}) on the James Clerk Maxwell Telescope (JCMT) during flexible observing in 2017 September 02,16,17,23,24, 2018 January 16-18, February 12 and March 29 (programs ID: M17BP024, M18AP054) under good weather conditions ($\tau_{225{\rm GHz}}\leq 0.06$). For each target, the observations covered a field-of-view of $\simeq13.7\arcmin$ in diameter using a {\sc daisy} scanning pattern centered on each ELAN position. Each field was scheduled with cycles (scans) of about 30 minutes, resulting in 3 hours (6 scans) for the MAMMOTH-1, the Slug and the Jackpot ELANe, while in 2.7 hours (5 scans) for the Fabulous ELAN. Even though SCUBA-2 acquires simultaneously data at 850 and 450~$\mu$m, here we focus only on the dataset at longer wavelengths as the 450~$\mu$m dataset is not deep enough to obtain proper number counts.
%it is the deepest. 

%MAMMOTH 6 scans of 30 min each -> 3 hours
%UM287   6 scans of 30 min each -> 3 hours
%Jackpot 6 scans of 30 min each -> 3 hours
%FAB are 5 scans of 32 min each -> 2.7 hours

The data reduction follows the method outlined in \citet{TC2013a} and \citet{FAB2018b}, which rely on the Dynamic Iterative Map Maker (DIMM) included in the SMURF package from the STARLINK software (\citealt{Jenness2011,Chapin2013}). Each scan has been reduced adopting the standard configuration file, {\it dimmconfig\_blank\_field.lis}, which is best for our science purposes. The coaddition of the reduced scans into final maps was then performed with the MOSAIC\_JCMT\_IMAGES recipe in PICARD, the Pipeline for Combining and Analyzing Reduced Data (\citealt{Jenness2008}).

\begin{figure*}[ht]
\centering
\includegraphics[scale = 0.47]{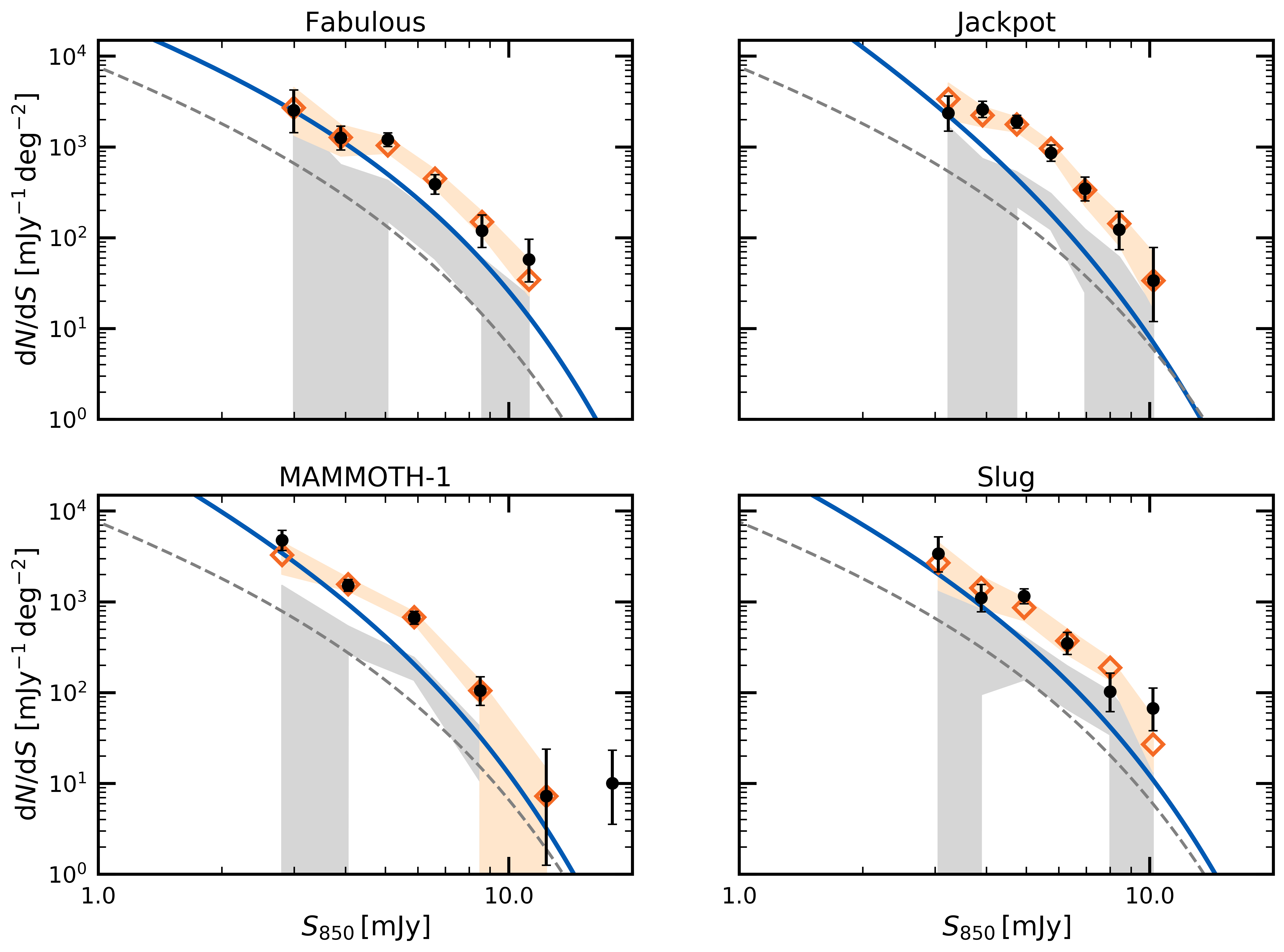}
\caption{
%Results of 500 realizations of the number counts models determined through Monte Carlo simulations (orange) and the fiducial model (gray). 
{Differential number counts recovered from 500 realizations of the number counts models determined through Monte Carlo simulations (orange) and the fiducial model (gray).} The shaded regions represent the $90\%$ confidence intervals of the counts recovered from the simulated maps. The black points are {raw} differential number counts, as defined in \autoref{sec:counts}. %\LEt{ Please capitalize the S in Section.}
The underlying counts models and the fiducial model (\citealt{Geach2017}) are plotted in blue and gray, respectively. The effect of flux boosting can be observed as the shaded regions are lying above their input model. {We note that the lower zero limits of the fiducial model realizations at faint fluxes reflect the non-uniform noise structure of our maps, where incompleteness increases with the distance, from the center of the map.}}
\label{fig:realizations}
\end{figure*}

% PREVIOUS CAPTION Results of 500 realizations of a fiducial model (gray curve; \citealt{Geach2017}) compared to pure source counts (black points). The red shaded regions represent the $90\%$ confidence intervals of the counts recovered from the simulated maps. This shows that in all four ELANe fields, the measured counts are significantly over abundant compared to those obtained in simulations by assuming the fiducial model.

Point source detectability is increased by applying a standard matched filter to these final maps. For this purpose, we used the PICARD recipe SCUBA2\_MATCHED\_FILTER. As a last step, we adopted the standard conversion factor for 850~$\mu$m (537 Jy pW$^{-1}$), with upwards of 10\% correction for flux calibration. {The uncertainty of the flux conversion factor at 850\,$\mu$m is typically 5\% \citep{Dempsey:2013aa}}

The final noise level at the centers of our maps (ELAN positions) is 0.88, 0.97, 1.02, 1.01~mJy~beam$^{-1}$, respectively, for the MAMMOTH-1, the Jackpot, the Slug, and the Fabulous ELANe. In this work, we focus on effective areas characterized by noise levels smaller than 2.5 times this central noise (Table~\ref{table:params} and white curves in Figure~\ref{fig:det}).

Furthermore, we produced a true noise map (jackknife map) for each field by subtracting two maps, each representing the sum of approximately half of that source's dataset. Therefore, the residual map should represent source-free noise data because all real sources should be subtracted irrespective of their significance. Importantly, we multiply these true noise maps by the scaling factor $\sqrt{t1\times t2}/(t1+t2), $ with t1 and t2 being the exposure time of each pixel from the two maps, to correct for the difference in exposure time. The central noise in these jackknife maps is consistent with the noise in the science data (0.88, 0.95, 1.03, 1.01~mJy~beam$^{-1}$, respectively, for MAMMOTH-1, the Jackpot, the Slug, and the Fabulous ELANe).

\section{Differential number counts}\label{sec:counts}

\begin{figure}[ht]
\centering
\includegraphics[scale = 0.37]{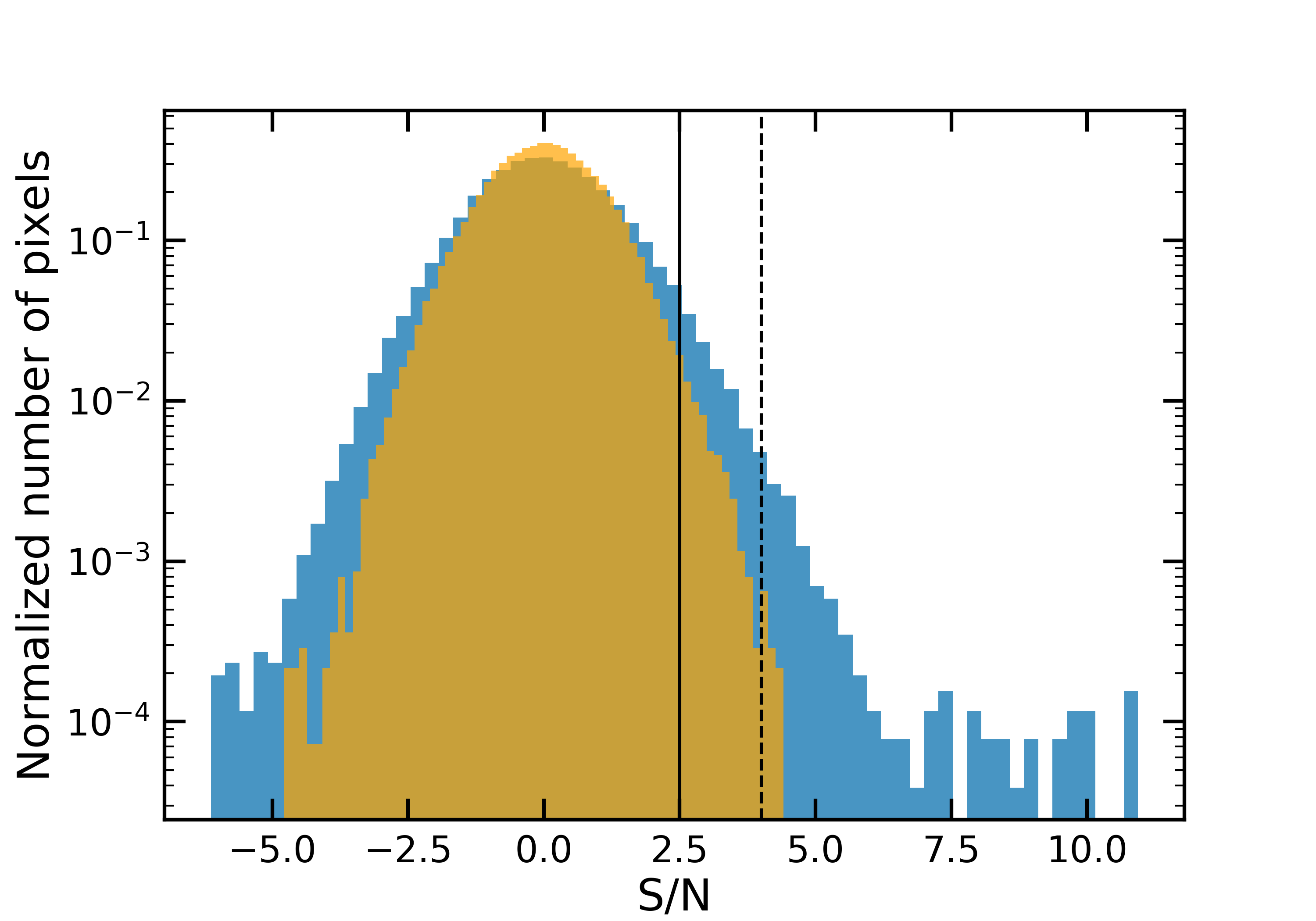}
\caption{{Histogram of signal-to-noise  ratio for pixels located within the effective area of the Jackpot map (blue) and the corresponding jackknife map (yellow). The solid vertical line represents the 2.5$\sigma$ cut used in our simulations. The dashed line represents the threshold of 4$\sigma$ above which we catalog sources.} {We note that the excessive negative signals compared to the jackknife map are a result of matched filtering, which can be properly accounted for during source extraction processes using a correct PSF.}}
\label{fig:histogram}
\end{figure}

{We extracted point sources from the SCUBA-2 maps} by following the procedure from \citealt{TC2013a}. The source extraction algorithm identifies the pixel with the highest signal-to-noise ratio (S/N) and subtracts a scaled point spread function (PSF) centered at the location of that pixel. The process continues until the highest S/N in the map falls below the detection threshold. In all four ELAN fields, we detect 73 sources with $>$ 4\,$\sigma$ located within effective areas of the maps. {The locations of the detected sources, plotted over (S/N) maps, are shown in Figure \ref{fig:det}, and their relevant information is listed in Tables A.1-A.4}
% Their detail information including coordinates and deboosted flux densities will be reported in a forthcoming paper.

To obtain the {raw} number counts, we first sum the number of sources in every flux bin and divide the sum by the width and detectable area of the bin. The detectable area is defined as the region of the map within which the sources with a given flux density can be detected above the S/N threshold. The flux bins are roughly in equal partition in logarithmic scale between the minimum and maximum flux densities of the detected sources. To correct for the detection of spurious sources, %counts obtained similarly from the jackknife maps 
{counts obtained from the jackknife maps using the same procedure} are subtracted off, and the results are the {raw} number counts. 
{The number of sources detected in the jackknife maps can be used to determine the fraction of spurious detections, which is 40\% at the simulation threshold of 2.5\,$\sigma$, and 7\% at the catalog threshold of 4 $\sigma$.} 
% \footnote{The fraction of spurious sources is about 40\% for a S/N of 2.5, which is the adopted threshold for our analyses.} 

The {raw} number counts are affected by observational biases, including flux boosting, {source blending}, and incompleteness. {Instead of making corrections for these biases separately, we chose to account for these effects altogether through Monte Carlo simulations.}
%To correct for these effects, we perform Monte Carlo simulations. 
{Our goal in the simulations is to determine underlying count models that correctly reproduce the {raw} number counts in every field}. We chose to model the differential number counts using a Schechter function of the form

% \begin{equation}
% \label{eqn:diff_counts}
%   \frac{dN}{dS} = \left\{
%   \begin{array}{l l}
%     {N_0}\left(\frac{S}{S_0}\right)^{-\alpha}  & \quad \text{ $S \leq S_0$}\\
%     {N_0}\left(\frac{S}{S_0}\right)^{-\beta}  & \quad \text{ $S > S_0$}\\
%   \end{array} \right.
% \end{equation} 

\begin{equation}
\label{eqn:schechter}
  \frac{dN}{dS} = \frac{N_0}{S_0} \left( \frac{S}{S_0}\right) ^{-\gamma}\exp\left(-\frac{S}{S_0}\right)
,\end{equation} 

where $N_0$ is the normalization and $S_0$ is the characteristic flux.

{The Schechter function or a double power-law function have been typically adopted for analyzing submillimeter number counts (e.g., \citealt{TC2013a,Geach2017}). Here, the Schechter function is preferred over a double power-law because of a smaller number of free parameters involved. It is also the model adopted by the fiducial blank-field counts and, thus, more straightforward comparisons can be made. Experimentally, we find  that the Schechter function reproduces the distribution of faint fluxes in our sample more accurately. We note, however, that the choice of the function does not significantly affect our results.}

Accurate positional information and low contamination rates are not critical for statistical analyses of number counts (\citealt{TC2013a}). To probe the faintest sources in our data set, we looked for the lowest detection threshold corresponding to statistically significant excess counts. {The adopted detection threshold of 2.5\,$\sigma$ is consistent with the excess in the positive signal, which is evident in Figure\ref{fig:histogram}.}

The simulation proceeds as follows. First, we fit the Schechter function to {raw} number counts using the least-squares method. The obtained initial parameters are used to populate the jackknife map with mock sources. {The positions of mock sources are selected randomly. The lowest fluxes we inject are 1 mJy, the same as that adopted by the fiducial blank-field model \citep{Geach2017} that we later use to compare our results.} At this value, the integrated flux density of our models is statistically consistent with the extragalactic background light (EBL; e.g., \citealt{Puget1996}). Our results are not, in fact, sensitive to the adopted flux levels of the faintest injected sources. For a given set of model parameters, we create 100 simulated maps and calculate the mean recovered counts. %The ratio of the {raw} counts to mean recovered counts is used to correct the input model before the next iteration. 
{The ratios of the {raw} counts to mean recovered counts, which are typically different in different flux bins, are used to create a new set of counts by multiplying the raw counts by these ratios. This new set of adjusted counts are then used for the next round of Schechter function fitting, for which the best fit is adopted as the new input model for the next iteration of the simulations.}
For each field, the process is repeated until the {convergence, which is when the }recovered counts fall within $1 \sigma$ error of the {raw} counts. 

The parameters of the underlying models obtained through this procedure are listed in Table \ref{table:params}. %Due to a high density of faint sources, our simulation is prone to overfitting. Specifically, an overestimation on the faint end can fully account for the bright sources via flux boosting. To avoid this effect, we keep the value of $S_0$ fixed and thus reduce the number of free parameters. 
Given the limited dynamical range of the measured counts, $N_0$ and $S_0$ tend to be degenerate. 
%As a result the simulations would not converge at times. 
We therefore keep the value of $S_0$ fixed to 2.5, same as the fiducial blank-field model \citep{Geach2017}, but the other two parameters free. %{We confirm that changing $S_0$ to other values does not significantly alter the results.} 
By doing so, we confirm that the simulations can converge and the results from each run are consistent with each other. {We also confirm that changing $S_0$ to other values does not significantly alter the results.} To verify our methods, we run the same analyses on the SCUBA-2 images of one known blank field, CDF-N, and we confirm that CDF-N shows no evidence of overdensity, which is consistent with the published results (\citealt{TC2013a}).

\begin{table*}
\begin{center}
\caption{Information of the four ELANe in our sample, along with their best-fit parameters for the underlying counts models formulated as Schecter functions.}
\begin{tabular}{ccccc|ccc}
\hline
\hline
ELAN Name  & R.A. & Decl. & Redshift & Effective area & N$_0$  &  S$_0$  &  $\gamma$  \\
      & (J2000; h:m:s) & (J2000; $^\circ$:$'$:$''$) & & (arcmin$^{-2}$)& (mJy$^{-1}$~deg$^{-2}$) & (mJy) &       \\
\hline
         Fabulous & 10:20:09.99 & +10:40:02.7  &3.17 & 106.7 & $27000 \pm 6400$ & 2.50 & $1.45 \pm 0.25$\\
         Jackpot & 08:41:59.26 & +39:21:40.0 &2.04 & 108.9 & $39900 \pm 7300$ & 2.50 & $2.68 \pm 0.24$\\
         MAMMOTH-1 & 14:41:27.62 & +40:03:31.4 &2.32 & 109.8 & $31200 \pm 4300$ & 2.50 & $2.12 \pm 0.18$\\
         Slug & 00:52:03.26 & +01:01:08.6 &2.28 & 109.5 &  $26200 \pm 10900$ & 2.50 & $2.05 \pm 0.41$\\
\hline
\hline
\end{tabular}
\label{table:params}
\end{center}
\end{table*}

%fab = counts_model(x, 27829.359008946707, 2.5000000000000004, 1.4272622961035133)
% slug = counts_model(x, 29946.138898957666,2.5000000000000004,1.9778855934478048)
% jackpot = counts_model(x,44522.694336434855, 2.5000000000000004, 2.4224951236229297)
% mammoth = counts_model(x, 37701.73274268243, 2.5000000000000004, 1.5825972939359536)

We tested whether the models determined through Monte Carlo simulations reproduce the {raw} number counts. We created 500 simulated maps for each ELAN field by injecting sources into the respective jackknife map according to the derived counts model. The mean recovered counts along with $90 \%$ confidence intervals are shown in Figure \ref{fig:realizations}. %We find a good agreement between the simulated counts and the counts from ELAN fields. 
{We find good agreements between the simulated counts and the raw counts in all four ELAN fields, confirming that our Monte Carlo simulations have properly accounted for observational biases.}
We note that the brightest bin in MAMMOTH-1 ELAN contains only two sources and its counts do not, in fact, differ substantially from the model; however, given the low probability of chance detection, they are often suspected to be associated with the ELAN system (\citealt{FAB2018b}).

%{Using the 500 simulated maps for every field, we estimate the flux boosting factors and positional uncertainties. In every map, we search for an injected source corresponding to a detected source. An injected source is considered a match if it is located within the beam area of a detected source. We define flux boosting as the ratio of the recovered source flux to the intrinsic source flux. The estimated flux boosting ratios are applied to correct flux densities of sources in our catalogs.}

To {further} verify that our procedure is able to constrain the overdensity, we {constructed another set of 500 simulated maps}
%repeat the {creation of 500 simulated maps} 
using a blank field model. {By injecting sources into the jackknife maps following a blank-field model, we simulated the number counts that would have been observed in each of the four fields if an overdensity were not present}. As a blank field model, we selected the Schechter function from \citet{Geach2017}, with the following parameters: $N_0 = 7180 \pm 1220$\,deg$^{-2}$, $S_0 = 2.5 \pm 0.4$ mJy, and $\gamma_0 = 1.5 \pm 0.4$, referred to as the fiducial model throughout this paper.  As shown in Figure \ref{fig:realizations}, the realizations of the blank field model result in recovered counts that are significantly lower than the {raw} counts from the ELAN fields, confirming the presence of overdensities. 

\section{Discussion} \label{sec:discussion}

\begin{figure*}[ht]
\centering
\includegraphics[width=0.9\textwidth]{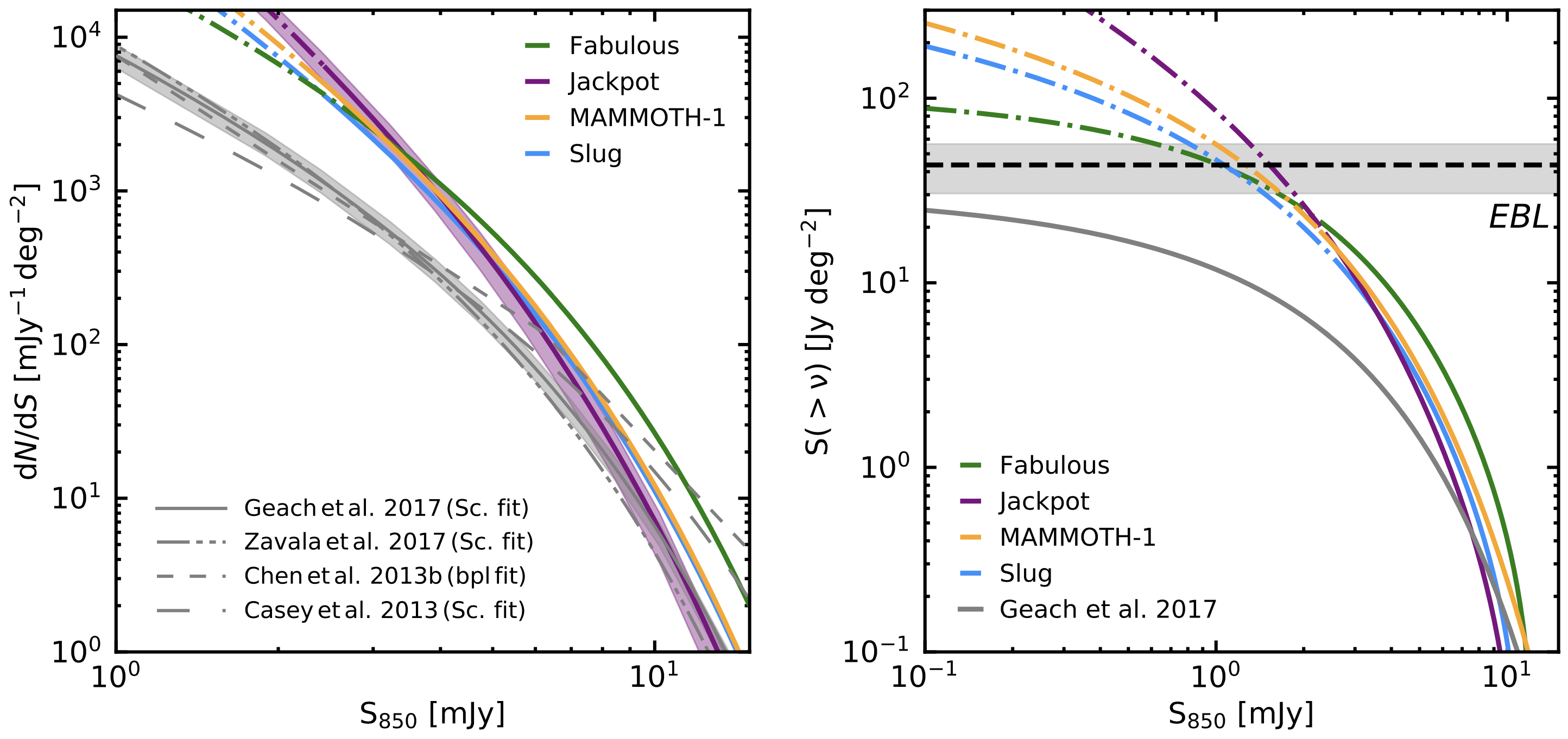}
\caption{%\LEt{ Please add a brief intro before the panel descriptions.} 
Best-fit number counts models and their integrated energy densities.
Left: Derived number counts models underlying distribution of sources in ELAN fields compared to blank field models (gray). {To highlight the typical uncertainties in these models, we show the confidence bands for the Geach et al. model and for the Jackpot field.} Right: Extragalactic background light as a function of flux. In both panels, the models for ELANe are plotted in solid at the flux range of the measured counts, and dashed in flux regimes fainter than our measurements.}
\label{fig:models}
\end{figure*}

In the previous section, we determine the number count models that reproduce the observed density of sources in ELAN fields. We plot these models in the left panel of Figure \ref{fig:models}, along with blank field models from the literature \citep{Chen:2013gq,Casey:2013aa,Geach2017,Zavala:2017aa}. A comparison between ELANe and blank fields reveals the presence of overdensities that extend from the lowest fluxes up to $6-7$ mJy. Above that threshold, the ELAN models become indistinguishable from the fiducial model based on the current statistics. We have tried splitting the counts by the distance to the ELAN, meaning the inner $<2'$ and outer $>2'$ regions, but we do not find any significant differences, so there is no evidence of counts overdensity as a function of radial distance, {%as it 
contrary to what was found in other submillimeter or millimeter counts studies toward other AGN samples or dense fields at high redshifts \citep{Zeballos:2018aa,Cooke:2019aa}.}

{As a test, we have tried different radial binning procedures but we found the results insensitive to these choices. There are a few factors that could be the root of the variety in the results compared to the previous work that suggested otherwise. First, there is the sample size. Previous studies have at least twice the number of fields compared to ours and some also have significantly deeper data. This is despite the fact that the observed differences in overdensity between the inner and outer regions reported in these studies are marginal, that is, close to or less than 3 sigma. We therefore expect that our current data quality would not have allowed us to make the detection if the SMGs around ELANe behave similarly to those in other dense fields. Second, while the first factor argues from a purely statistical point of view, it is also possible that the SMGs intrinsically span a wider physical range in the environments around ELANe compared to those around HzRGs or X-ray selected clusters targeted by other studies. The fact that the footprint of our SCUBA-2 observations (\autoref{fig:det}) matches the expected span ($\sim$10-20 cMpc) of a protocluster hosting a massive (10$^{13}$-10$^{14}$\,M$_\odot$) halo at the redshift of our sample ELANe (Chiang et al. 2013) makes the scenario of more widespread SMG distribution around ELANe a plausible hypothesis.}

To %quantify and estimate 
quantify the overdensity factor for each of the ELANe, we adopted two approaches. In the first approach, which was also adopted in other studies \citep{FAB2018a}, %we first correct the {raw} counts for flux boosting. 
{for each field we first correct the raw counts for observational biases by dividing the raw counts in each flux bin by the ratios of the mean recovered counts to its underlying model obtained from the simulations.}
We then fit the fiducial model to these corrected counts, allowing only the normalization $N_0$ to vary. The overdensity factor is defined as the ratio of the obtained normalization to fiducial model normalization. As a result, we find that the overdensity factors for each field are $2.7 \pm 0.5$, $2.0 \pm 0.4$, $1.8 \pm 0.5$, $3.8 \pm 0.7,$ respectively. for MAMMOTH-1, the Jackpot, the Slug, and the Fabulous ELANe. The uncertainties in each field have taken into account both Poisson noises and those from the simulations. The weighted average overdensity factor in our sample is $2.3 \pm 0.2$\footnote{If we instead adopt 0.1\,mJy as the faintest injected sources for the simulations, the factor would be $2.2+/-0.3$, consistent to within 1\,$\sigma$ to the adopted method.}. 

{The first approach is appropriate if the shape of the model in ELANe agrees with that of the fiducial model. However, as shown in \autoref{fig:realizations} and \autoref{table:params} there is tentative evidence that the models in ELANe are steeper, so the overdensity could be flux dependent. In such cases, estimating overdensity factors by only fitting the normalization would bias the results to the lower end. To mitigate this bias we therefore estimate the overdensity factors by simply %contrasting 
comparing the cumulative number of sources between each ELAN field and the fiducial field. The cumulative number of sources are obtained by integrating the models within the flux range probed by our SCUBA-2 observations ($S_{850}\sim2.5-10\,{\rm mJy}$). As a result, we find overdensity factors of $3.6 \pm 0.9$, $3.8 \pm 1.2$, $2.9 \pm 1.8$, and $3.9 \pm 1.5$ for MAMMOTH-1, the Jackpot, the Slug, and the Fabulous ELANe, respectively. These give a weighted overdensity factor of $3.6\pm0.6$\footnote{3.5$\pm$0.7 if we instead adopt 0.1\,mJy as the faintest injected sources for the simulations.}. {As expected, the overdensity factors using the cumulative approach are slightly larger than the normalization approach in the three fields where the best-fit underlying models have slightly steeper slopes. In addition, since the cumulative method additionally includes the uncertainties from the bright-end slope, $\gamma$, the uncertainties of the overdensity factors would be larger compared to those of the first approach with fittings only to the normalization. The results from both approaches however are consistent with each other}, thus we chose to adopt this second approach since it does not suffer from any possible bias due to the assumptions of the model shape.} 
%Note the error-normalized standard deviation based on the two approaches are both close to one, meaning the variation of the overdensity factor is mainly due to measurement uncertainties and there is no evidence of intrinsic scatter.
 
Our results are consistent with those found in our pilot study on the MAMMOTH-1 ELAN alone \citep{FAB2018b}. By expanding the measurements to include three more ELANe, we have found {overabundant submillimeter sources around} all four ELANe and obtained a more precise average overdensity factor. With these results, we conclude that there are significantly (with a $>4\,\sigma$ confidence level) more SMGs around ELANe in general. We postulate that this overdensity is likely caused by additional SMGs that are physically associated with each ELAN, but future emission line measurements are needed to confirm this hypothesis; this would also allow us to measure the three-dimensional overdensity factors. Our results are also in agreement with submillimeter and millimeter surveys in other samples of AGN, such as high-redshift radio galaxies (HzRGs) and {\it WISE}-selected AGN, in which (on average) a factor of 2-6 in overdensity was found (e.g., \citealt{Rigby:2014aa,Jones:2017aa,Zeballos:2018aa}). {What we found to be different is that, while it comes with a varying range of significance, the overabundant submillimeter sources is found in all the surveyed ELAN fields;  this ubiquity was not observed in other studies where typically less than half of the targeted fields show any signs of overdensity. While surveying more ELAN fields is needed,}
%{What is different is that we found evidence of overdensity in all the surveyed fields, and this ubiquity was not observed in other studies.} 
%These 
these findings support the fact that certain levels of co-evolution exist between SMGs and AGN, as they are likely part of similarly dense environments, {and the co-evolution may be more commonly observed in the ELAN fields.}

{On the other hand,} if our best-fit count models continue to the fainter end, our findings could have implications for the extragalactic background light (EBL) measurements at 850\,$\mu$m. In the right panel of Figure \ref{fig:models}, we plot the integrated flux density above a given flux density based on the best-fit models for each ELAN field as well as the fiducial model. By comparing the results to the background light measurements from the COBE satellite (Fixen et al. 1996), we find that in the ELAN fields, the background light can already be fully accounted for by integrating the counts models down to $1-2$\,mJy, and there would be significantly excessive background light if the models are integrated down to 0.1\,mJy level -- which is the  faintest detection level reported for other fields (e.g., \citealt{Chen:2014aa,Fujimoto:2016aa,Hsu:2017aa,Gonzalez-Lopez:2017ab,Gonzalez-Lopez:2020aa}). Our results therefore suggest that there could be up to an order of magnitude in fluctuations with regard to the 850\,$\mu$m background on the survey scales toward these dense fields. {On the other hand, if this level of fluctuation does exist, we would expect these extreme fluctuations to be rare; otherwise, a deficit of background light at a similar level would have to exist in some other fields to bring the average down to the observed level by COBE.} {Deeper observations are key for constructing the shape of the number count distribution at the faint end, which, in turn, will provide better constraints in this area.}
%{Deeper observations which allow constructions of the fainter end counts shape would provide tighter constraints on this issue.}

\begin{figure}[ht]
\centering
\includegraphics[scale = 0.8]{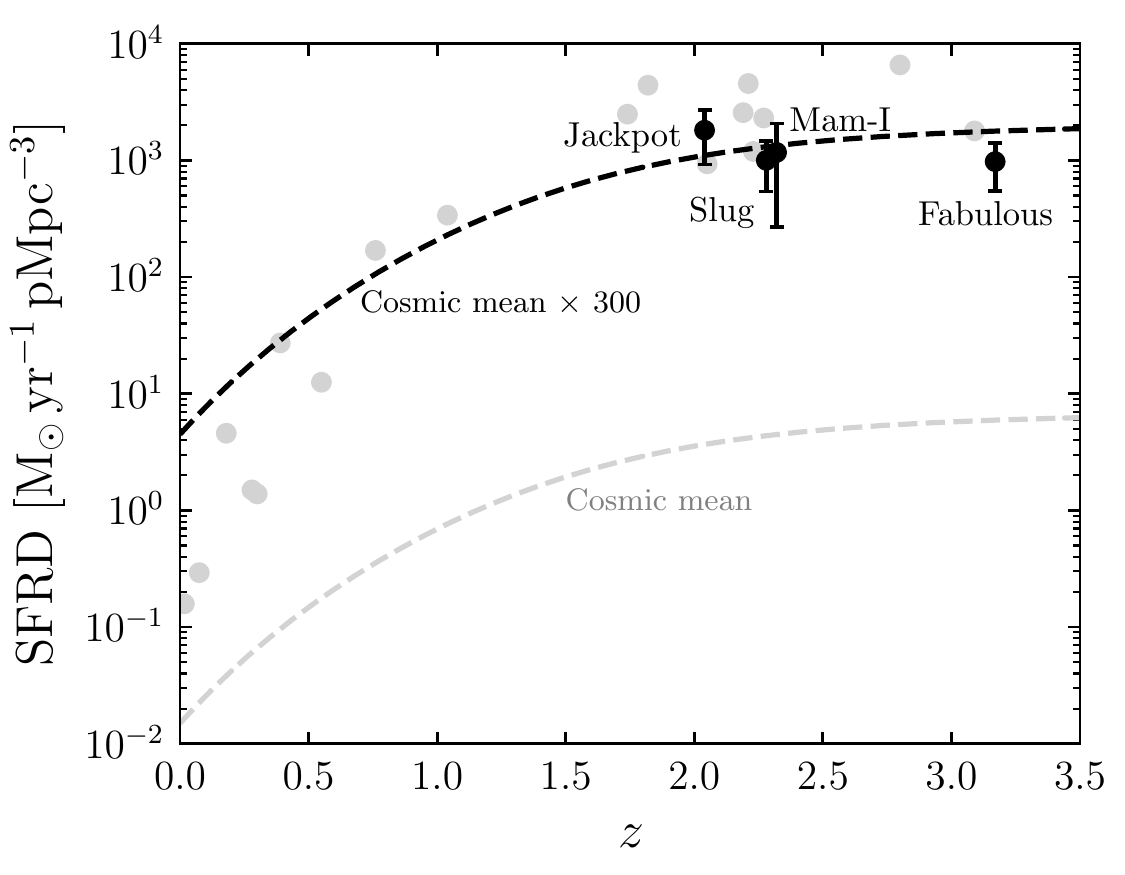}
\caption{{Cosmic SFR density in physical volumes. The estimates based on our counts measurements are plotted in black points, with the corresponding field names indicated.
%, and 
The gray points mark the values reported in the literature around samples of overdense regions \citep{Clements:2014aa,Kato:2016aa} {that are not known to host ELANe.} The cosmic mean is plotted in gray dashed based on eqn. 15 of \citet{Madau:2014aa} but converted to physical volume density, 
%and dashed 
whilst the black curve shows the cosmic mean multiplied by a factor of 300.}}
\label{fig:sfrd}
\end{figure}

Finally, we attempted to estimate the total obscured star formation within the Mpc scale environments of ELAN, which can be achieved by assuming that the excess number of submillimeter sources are all associated with the respective ELAN. To compute the SFR densities we first seek a proper conversion from $S_{\rm 850}$ to SFRs. We utilized a sample of ALMA-identified SMGs in the UKIDSS-UDS field, which has been studied in detail with proper SED fittings \citep{Ugne2020}. {Similarly to our observations, the sample of \citet{Ugne2020} was drawn from a flux-limited sample produced by the SCUBA-2 Cosmology Legacy Survey \citep{Geach2017}. Their results should therefore be representative, on average, of SCUBA-2 sources uncovered in other fields.} These authors found a linear correlation of log$_{10}$[SFR(M$_\odot$yr$^{-1}$)] = (0.42$\pm$0.06)$\times$log$_{10}$[$S_{\rm 870}$(mJy)]+(2.19$\pm$0.03) for their SMGs, which span a flux range of $\sim1-10$\,mJy, appropriate for the SCUBA-2 sources discovered in our target ELAN fields.
%We perform a linear fit to the SMGs that have $S_{\rm 850}$=1-10\,mJy and are located at the same redshift range as our ELAN fields and find log$_{10}$(SFR) = (0.5$\pm$0.2)$\times$log$_{10}$($S_{\rm 850}$)+(2.1$\pm$0.1) with a scatter of $\sim$0.3\,dex. 
We then computed the total SFR densities by integrating over a given $S_{\rm 850}$ range in which the SFR contribution at a given $S_{\rm 850}$ is the product of the excess number density of the submillimeter sources and their corresponding SFR based on the conversion. For each field, by considering a flux range of $S_{\rm 850}=1-20$\,mJy and assuming a sphere with an equivalent circular radius of the corresponding effective area, we obtain SFR densities of 1100$\pm$500, 1100$\pm$500, 2300$\pm$1100, and 1400$\pm$1100\,$M_\odot$\,yr$^{-1}$\,Mpc$^{-3}$ for Fabulous, Slug, Jackpot, and MAMMOTH-1 ELAN, and a weighted average SFR density of $\Sigma$SFR= 1200$\pm$300\,$M_\odot$\,yr$^{-1}$\,Mpc$^{-3}$. {We plot the results in \autoref{fig:sfrd}, showing that they are consistent with those found in the Mpc-scale environments of other quasar samples or proto-clusters at $z\sim2-3$ \citep{Clements:2014aa,Dannerbauer2014,Kato:2016aa}. This result suggests that the star formation activities around ELANe are at a similar level of other dense systems in this redshift range, or, in other words, at a factor of about 300 greater than the cosmic mean.} 

\section{Summary} \label{sec:summary}
{In this second paper of the AMUSE$^2$ series, we present the results of a SCUBA-2 850\,$\mu$m survey around four ELANe, with the aim of understanding the dust-obscured star formation around these massive systems on Mpc scales. We summarize our findings in the following.

\begin{enumerate}
\item By carefully performing the number counts analyses through self-consistent simulations, we find that the number of submillimeter sources with $S_{\rm 850}\gtrsim3$\,mJy are overabundant in all four ELANe fields, by a factor of 2-4 within $\sim$5\,Mpc. The significance of overdensity detection in each field varies between 1-4\,$\sigma$, depending on the precise methods used in computing these overdensity factors. The weighted average overdensity factor is 2.3$\pm$0.2, if normalized to the fiducial blank-field counts, or $3.6\pm0.6,$ if instead adopting a cumulative method over the flux density range probed.

\item By integrating the number count models in the ELAN fields, we find that the 850\,$\mu$m EBL would be fully accounted for at 1-2\,mJy, and up to an order of magnitude higher when integrating down to about 0.1\,mJy; thereby suggesting an order of magnitude fluctuations of the 850\,$\mu$m EBL on the survey scales toward these dense fields. Deeper observations are key to constraining the count shapes at the faint end, which, in turn, will provide better constraints in this regard.

\item Finally, assuming that all the excessive sources can be attributed to SMGs associated with the corresponding ELANe, and by adopting a linear $S_{\rm 850}$-to-SFR relation reported by a general SMG study \citep{Ugne2020}, we find SFR densities of 1000-2000\,M$_\odot$ yr$^{-1}$ Mpc$^{-3}$ with a weighted average of $\Sigma$SFR=$1200\pm300$\,M$_\odot$ yr$^{-1}$ Mpc$^{-3}$, which is a factor of about 300 larger than the cosmic mean. These are consistent with those found in other quasar samples or proto-clusters, suggesting that the presence of ELANe in the central $\sim$100\,kpc regions does not significantly affect the dust-obscured star-forming activities on Mpc scales.
%\LEt{ Please add 2-3 sentences of summary as a final paragraph after the bullets.}
\end{enumerate}
This work represents the first step into systematically quantifying the Mpc-scale environments around ELAN from the perspective of dusty star formation. Follow up observations with spectroscopic measurements to confirm member associations would be the next step to understand their spatial and kinematic distributions, key to unlock the formation processes on $z\sim2-3$ massive proto-clusters. 

} 

%suggesting that the star formation activities around ELANe are at a similar level of other dense systems in this redshift range, which is a factor of about 300 larger than the cosmic mean.}
%By considering the flux range of our observations ($S_{\rm 850}>3$\,mJy) and the effective area, we obtain an average SFR density of log$_{10}$($\Sigma$SFR)=$-1.3$\,$M_\odot$\,yr$^{-1}$\,Mpc$^3$, $\sim$40\% of the cosmic mean (Madau \& Dickinson 2014). If we instead integrate down to 1\,mJy the SFR density would be about log$_{10}$($\Sigma$SFR)=$-0.4$\,$M_\odot$\,yr$^{-1}$\,Mpc$^3$, a factor of three above the cosmic mean. 

%representing the extragalactic background light (e.g. Puget et al. 1996) (as a function of flux) . Already by $1-2$ mJy the calculated EBL values exceed the background light measurements from the COBE satellite (Fixen et al. 1996).
%To match the fiducial EBL value, the ELAN flux density models would need to eventually converge with blank field predictions at low fluxes. At the current depth of our sample, we are unable to explore this flux regime. 

\begin{acknowledgements}
      We thank the first ever ESO summer research program \citep{2019Msngr.178...57M} that facilitated the start of this research project; Ian Smail for providing useful comments on this work; and the reviewer for a thoughtful report that has improved the manuscript. C.C.C. acknowledges support from the Ministry of Science and Technology of Taiwan (MOST 109-2112-M-001-016-MY3). MN and FAB warmly thank the JCMT staff for their hospitality and support during the observatory visits and observing runs. This project has received funding from the European Research Council (ERC) under the European Union's Horizon 2020 research and innovation programme (grant agreement No 757535). Y.Y.'s research was supported by Basic Science Research Program through the National Research Foundation of Korea (NRF) funded by the Ministry of Science, ICT \& Future Planning (NRF-2019R1A2C4069803). The James Clerk Maxwell Telescope is operated by the East Asian Observatory on behalf of The National Astronomical Observatory of Japan; Academia Sinica Institute of Astronomy and Astrophysics; the Korea Astronomy and Space Science Institute; Center for Astronomical Mega-Science (as well as the National Key R\&D Program of China with No. 2017YFA0402700). Additional funding support is provided by the Science and Technology Facilities Council of the United Kingdom and participating universities and organizations in the United Kingdom and Canada. Additional funds for the construction of SCUBA-2 were provided by the Canada Foundation for Innovation. {This research made use of Astropy,\footnote{http://www.astropy.org} a community-developed core Python package for Astronomy \citep{astropy:2013, astropy:2018}.} The authors wish to recognize and acknowledge the very significant cultural role and reverence that the summit of Maunakea has always had within the indigenous Hawaiian community. We are most fortunate to have the opportunity to conduct observations from this mountain. 
\end{acknowledgements}

\bibliographystyle{aa} 
\bibliography{AMUSE2.bib}

\begin{appendix} 
\section{Source catalogs}
In this appendix, we provide the source catalogs for $\ge$4\,$\sigma$ detection. The last two columns of each table represent the deboosted flux densities ($f_{850}^{\rm Deboosted}$) and the positional uncertainties ($\Delta(\alpha,\delta)$){, which are estimated as follows. We use the 500 simulated maps described in \autoref{sec:counts} %\LEt{ uppercase S.}
in every field to estimate the flux boosting factors and positional uncertainties. 
%for the sources that are detected with a statistical significance above 4\,$\sigma$.
%for the significantly ($\ge$4\,$\sigma$) sources. 
In every map, we search for an injected source corresponding to a detected source. An injected source is considered a match if it is located within the beam area of a detected source. We define flux boosting as the ratio of the recovered source flux to the intrinsic source flux. Both flux boosting and positional offsets are recorded for each injected source and we find they are both a function of S/N, which is consistent with previous findings (e.g., \citealt{Geach2017}). In the catalogs, we therefore apply the median flux boosting correction for each source given its S/N and we quote the corresponding positional uncertainty.
%The estimated flux boosting ratios are applied to correct source flux densities in our catalogs, which are presented in the Appendix.
}
%, estimated based on similations described in \autoref{sec:counts}.

\begin{table*}[ht!]
\begin{center}
\caption{SCUBA-2 850\,$\mu$m detected sources within the assumed effective area around the Slug ELAN}
\begin{tabular}{ccccccc}
\hline
\hline
Name & R.A.  & Decl. & S/N & $f_{850}$ & $f_{850}^{\rm Deboosted}$ & $\Delta(\alpha,\delta)$ \\
 & (J2000; h:m:s) & (J2000; $^\circ$:$'$:$''$) & & (mJy)& (mJy)& arcsec \\  
\hline
Slug-850.1&00:52:01.6&+01:03:41.2&8.7&$11.4 \pm 1.3$&$9.6 \pm 1.1$&0.8\\
Slug-850.2&00:51:54.4&+01:05:31.2&5.4&$10.1 \pm 1.9$&$7.1 \pm 1.3$&1.6\\
Slug-850.3&00:52:16.4&+00:58:57.2&5.1&$9.0 \pm 1.8$&$6.1 \pm 1.2$&1.6\\
Slug-850.4&00:52:20.5&+01:01:07.2&4.9&$8.9 \pm 1.8$&$5.8 \pm 1.2$&1.8\\
Slug-850.5&00:52:00.9&+01:01:55.2&4.4&$4.6 \pm 1.0$&$2.7 \pm 0.6$&2.0\\
Slug-850.6&00:51:54.5&+01:01:23.2&4.3&$5.0 \pm 1.2$&$2.9 \pm 0.7$&2.0\\
Slug-850.7&00:52:15.2&+01:05:31.2&4.2&$9.0 \pm 2.1$&$5.0 \pm 1.2$&2.2\\

\hline
\hline
\end{tabular}
\end{center}
\end{table*}

\begin{table*}[]
\begin{center}
\caption{SCUBA-2 850\,$\mu$m detected sources within the assumed effective area around the Jackpot ELAN}
\begin{tabular}{ccccccc}
\hline
\hline
Name & R.A.  & Decl. & S/N & $f_{850}$ & $f_{850}^{\rm Deboosted}$ & $\Delta(\alpha,\delta)$ \\
 & (J2000; h:m:s) & (J2000; $^\circ$:$'$:$''$) & & (mJy)& (mJy)& arcsec \\  
\hline
Jackpot-850.1&08:41:58.8&+39:21:51.0&10.9&$11.0 \pm 1.0$&$9.0 \pm 0.8$&0.9\\
Jackpot-850.2&08:42:06.9&+39:20:37.0&6.0&$6.4 \pm 1.1$&$3.8 \pm 0.6$&1.8\\
Jackpot-850.3&08:42:16.9&+39:19:28.9&5.5&$7.9 \pm 1.4$&$4.4 \pm 0.8$&2.0\\
Jackpot-850.4&08:42:14.7&+39:23:12.9&5.2&$7.5 \pm 1.4$&$4.0 \pm 0.8$&2.0\\
Jackpot-850.5&08:41:46.6&+39:20:07.0&5.1&$8.0 \pm 1.6$&$4.3 \pm 0.8$&2.0\\
Jackpot-850.6&08:42:19.2&+39:18:50.9&5.0&$7.8 \pm 1.6$&$4.0 \pm 0.8$&2.2\\
Jackpot-850.7&08:42:03.3&+39:17:33.0&5.0&$7.7 \pm 1.6$&$4.0 \pm 0.8$&2.2\\
Jackpot-850.8&08:42:00.5&+39:18:45.0&4.9&$6.6 \pm 1.3$&$3.4 \pm 0.7$&2.2\\
Jackpot-850.9&08:42:05.0&+39:19:57.0&4.6&$5.1 \pm 1.1$&$2.5 \pm 0.5$&2.2\\
Jackpot-850.10&08:42:11.9&+39:26:35.0&4.6&$9.7 \pm 2.1$&$4.8 \pm 1.0$&2.2\\
Jackpot-850.11&08:41:59.2&+39:20:43.0&4.6&$4.6 \pm 1.0$&$2.3 \pm 0.5$&2.2\\
Jackpot-850.12&08:42:19.0&+39:21:04.9&4.6&$6.4 \pm 1.4$&$3.2 \pm 0.7$&2.2\\
Jackpot-850.13&08:42:09.0&+39:21:03.0&4.6&$5.1 \pm 1.1$&$2.5 \pm 0.6$&2.2\\
Jackpot-850.14&08:42:00.7&+39:25:15.0&4.5&$6.6 \pm 1.5$&$3.2 \pm 0.7$&2.3\\
Jackpot-850.15&08:42:23.1&+39:23:00.8&4.4&$6.9 \pm 1.6$&$3.3 \pm 0.8$&2.3\\
Jackpot-850.16&08:42:25.5&+39:21:26.8&4.3&$7.0 \pm 1.6$&$3.4 \pm 0.8$&2.3\\
Jackpot-850.17&08:42:10.7&+39:19:19.0&4.3&$5.9 \pm 1.4$&$2.9 \pm 0.7$&2.3\\
Jackpot-850.18&08:42:06.2&+39:17:59.0&4.3&$6.9 \pm 1.6$&$3.2 \pm 0.8$&2.4\\
Jackpot-850.19&08:42:00.5&+39:20:53.0&4.2&$4.1 \pm 1.0$&$1.9 \pm 0.5$&2.4\\
Jackpot-850.20&08:42:02.3&+39:19:49.0&4.2&$4.5 \pm 1.1$&$2.1 \pm 0.5$&2.4\\
Jackpot-850.21&08:41:53.5&+39:21:47.0&4.1&$4.4 \pm 1.1$&$2.0 \pm 0.5$&2.4\\
Jackpot-850.22&08:42:05.7&+39:18:35.0&4.0&$5.8 \pm 1.5$&$2.7 \pm 0.7$&2.4\\
Jackpot-850.23&08:42:00.5&+39:16:11.0&4.0&$7.0 \pm 1.8$&$3.3 \pm 0.8$&2.4\\
Jackpot-850.24&08:41:51.7&+39:22:43.0&4.0&$5.0 \pm 1.2$&$2.3 \pm 0.6$&2.4\\

\hline
\hline
\end{tabular}
\end{center}
\end{table*}\label{tabA2}

\begin{table*}[]
\begin{center}
\caption{SCUBA-2 850\,$\mu$m detected sources within the assumed effective area around the Fabulous ELAN}
\begin{tabular}{ccccccc}
\hline
\hline
Name & R.A.  & Decl. & S/N & $f_{850}$ & $f_{850}^{\rm Deboosted}$ & $\Delta(\alpha,\delta)$ \\
 & (J2000; h:m:s) & (J2000; $^\circ$:$'$:$''$) & & (mJy)& (mJy)& arcsec \\  
\hline
Fabulous-850.1&10:20:10.0&+10:40:08.0&11.1&$11.3 \pm 1.0$&$9.9 \pm 0.9$&0.5\\
Fabulous-850.2&10:20:19.5&+10:36:40.0&8.9&$12.0 \pm 1.3$&$10.4 \pm 1.2$&0.7\\
Fabulous-850.3&10:20:28.9&+10:39:34.0&7.9&$12.6 \pm 1.6$&$10.5 \pm 1.3$&0.8\\
Fabulous-850.4&10:20:27.1&+10:41:54.0&5.8&$10.4 \pm 1.8$&$7.9 \pm 1.4$&1.4\\
Fabulous-850.5&10:20:23.3&+10:40:10.0&5.1&$7.0 \pm 1.4$&$5.1 \pm 1.0$&1.6\\
Fabulous-850.6&10:20:26.3&+10:41:36.0&5.1&$8.2 \pm 1.6$&$5.8 \pm 1.2$&1.7\\
Fabulous-850.7&10:20:08.5&+10:38:52.0&5.0&$5.2 \pm 1.0$&$3.7 \pm 0.7$&1.7\\
Fabulous-850.8&10:20:22.2&+10:40:48.0&5.0&$6.5 \pm 1.3$&$4.6 \pm 0.9$&1.7\\
Fabulous-850.9&10:20:25.2&+10:41:18.0&4.9&$7.3 \pm 1.5$&$5.2 \pm 1.1$&1.7\\
Fabulous-850.10&10:19:58.7&+10:36:46.0&4.8&$8.7 \pm 1.8$&$6.2 \pm 1.3$&1.7\\
Fabulous-850.11&10:20:02.5&+10:37:16.0&4.7&$8.0 \pm 1.7$&$5.5 \pm 1.2$&1.8\\
Fabulous-850.12&10:20:17.5&+10:36:46.0&4.7&$5.9 \pm 1.3$&$4.1 \pm 0.9$&1.8\\
Fabulous-850.13&10:20:03.4&+10:39:14.0&4.6&$5.2 \pm 1.1$&$3.6 \pm 0.8$&1.8\\
Fabulous-850.14&10:20:08.8&+10:39:22.0&4.4&$4.5 \pm 1.0$&$3.0 \pm 0.7$&1.9\\
Fabulous-850.15&10:20:06.7&+10:35:10.0&4.3&$7.7 \pm 1.8$&$5.1 \pm 1.2$&1.9\\
Fabulous-850.16&10:20:06.6&+10:34:28.0&4.3&$8.3 \pm 1.9$&$5.5 \pm 1.3$&1.9\\
Fabulous-850.17&10:19:51.5&+10:42:18.0&4.3&$7.4 \pm 1.7$&$4.8 \pm 1.1$&2.0\\
Fabulous-850.18&10:19:52.1&+10:36:18.0&4.3&$9.7 \pm 2.3$&$6.3 \pm 1.5$&2.0\\
Fabulous-850.19&10:20:10.1&+10:40:00.0&4.0&$4.1 \pm 1.0$&$2.6 \pm 0.7$&2.0\\

\hline
\hline
\end{tabular}
\end{center}
\end{table*}\label{tabA3}

\begin{table*}[]
\begin{center}
\caption{SCUBA-2 850\,$\mu$m detected sources within the assumed effective area around the MAMMOTH-1 ELAN}
\begin{tabular}{ccccccc}
\hline
\hline
Name & R.A.  & Decl. & S/N & $f_{850}$ & $f_{850}^{\rm Deboosted}$ & $\Delta(\alpha,\delta)$ \\
 & (J2000; h:m:s) & (J2000; $^\circ$:$'$:$''$) & & (mJy)& (mJy)& arcsec \\  
\hline
MAMMOTH-850.1&14:41:25.5&+40:00:31.4&16.0&$21.0 \pm 1.3$&$18.3 \pm 1.1$&0.6\\
MAMMOTH-850.2&14:41:29.0&+40:01:19.4&15.9&$18.8 \pm 1.2$&$16.4 \pm 1.0$&0.6\\
MAMMOTH-850.3&14:41:45.6&+40:08:13.3&7.7&$13.9 \pm 1.8$&$10.9 \pm 1.4$&1.1\\
MAMMOTH-850.4&14:41:40.2&+40:01:01.4&7.4&$10.0 \pm 1.3$&$7.8 \pm 1.1$&1.1\\
MAMMOTH-850.5&14:41:44.7&+40:02:19.3&6.9&$8.9 \pm 1.3$&$6.7 \pm 1.0$&1.3\\
MAMMOTH-850.6&14:41:25.5&+40:07:29.4&6.2&$8.1 \pm 1.3$&$5.8 \pm 0.9$&1.5\\
MAMMOTH-850.7&14:41:14.9&+40:07:59.4&5.9&$9.3 \pm 1.6$&$6.3 \pm 1.1$&1.5\\
MAMMOTH-850.8&14:41:47.5&+40:00:01.3&5.4&$8.5 \pm 1.6$&$5.8 \pm 1.1$&1.8\\
MAMMOTH-850.9&14:41:35.6&+39:59:27.4&5.2&$7.6 \pm 1.5$&$4.8 \pm 0.9$&1.8\\
MAMMOTH-850.10&14:41:15.2&+40:06:01.4&5.1&$7.2 \pm 1.4$&$4.5 \pm 0.9$&1.8\\
MAMMOTH-850.11&14:41:49.4&+40:01:27.3&5.1&$7.3 \pm 1.4$&$4.6 \pm 0.9$&1.8\\
MAMMOTH-850.12&14:41:30.8&+40:03:07.4&4.9&$4.5 \pm 0.9$&$2.8 \pm 0.6$&1.8\\
MAMMOTH-850.13&14:41:46.8&+40:05:29.3&4.9&$6.7 \pm 1.4$&$4.2 \pm 0.9$&1.8\\
MAMMOTH-850.14&14:41:24.5&+40:03:07.4&4.9&$4.6 \pm 0.9$&$2.9 \pm 0.6$&1.8\\
MAMMOTH-850.15&14:41:15.3&+40:01:41.4&4.7&$6.7 \pm 1.4$&$3.9 \pm 0.8$&2.1\\
MAMMOTH-850.16&14:41:33.9&+40:01:41.4&4.5&$5.2 \pm 1.2$&$3.0 \pm 0.7$&2.1\\
MAMMOTH-850.17&14:41:35.6&+40:06:09.4&4.2&$5.2 \pm 1.2$&$3.0 \pm 0.7$&2.1\\
MAMMOTH-850.18&14:41:29.9&+40:07:43.4&4.2&$5.4 \pm 1.3$&$3.1 \pm 0.7$&2.1\\
MAMMOTH-850.19&14:41:44.5&+40:01:23.3&4.2&$5.7 \pm 1.4$&$3.3 \pm 0.8$&2.1\\
MAMMOTH-850.20&14:41:25.0&+40:06:29.4&4.1&$5.2 \pm 1.3$&$3.0 \pm 0.7$&2.1\\
MAMMOTH-850.21&14:41:44.2&+40:00:49.3&4.1&$5.7 \pm 1.4$&$3.3 \pm 0.8$&2.1\\ %extra one
MAMMOTH-850.22&14:41:28.7&+39:59:31.4&4.0&$5.7 \pm 1.4$&$3.3 \pm 0.8$&2.1\\
MAMMOTH-850.23&14:41:17.7&+39:58:09.4&4.0&$8.1 \pm 2.0$&$4.7 \pm 1.2$&2.1\\

\hline
\hline
\end{tabular}
\end{center}
\end{table*}\label{tabA4}

\end{appendix}
\end{document}